\documentclass[aps,prl,twocolumn, superscriptaddress, longbibliography]{revtex4-2} 
\usepackage{bm}% bold math
\usepackage[T1]{fontenc} 
\usepackage[utf8]{inputenc} 
\usepackage{lmodern} 
\usepackage{amsmath,amssymb,bm} 
\usepackage{graphicx}
\usepackage{siunitx} 
\usepackage{dcolumn} % align table columns on decimal point 
\usepackage{hyperref} 
\usepackage{microtype} 
\hypersetup{colorlinks=true, linkcolor=blue, citecolor=blue, urlcolor=blue}
\begin{document}

\title{Polymeric Solvents Control Swelling-Induced Surface Creasing}
\author{Zechao Jiang}
\affiliation{School of Physics, Beihang University, Beijing 102206, China}
\author{Zhaoyu Ding}
\affiliation{School of Physics, Beihang University, Beijing 102206, China}
\author{Shaohua Yang}
\affiliation{School of Mechanical Engineering and Automation, Beihang University, Beijing 102206, China}
\author{Ye Xu}
\affiliation{School of Mechanical Engineering and Automation, Beihang University, Beijing 102206, China}
\affiliation{Hangzhou International Innovation Institute, Beihang University, Hangzhou 311115, China}
\author{Dongshi Guan}
\affiliation{State Key Laboratory of Nonlinear Mechanics, Institute of Mechanics, Chinese Academy of Sciences, Beijing 100190, China}
\affiliation{School of Engineering Science, University of Chinese Academy of Sciences, Beijing 100049, China}

\author{Abdelhamid Maali}
\affiliation{Univ. Bordeaux, CNRS, LOMA, UMR 5798, 33405 Talence, France}

\author{Joshua D. McGraw}
\affiliation{Gulliver UMR 7083 CNRS, PSL Research University, ESPCI Paris, 10 rue Vauquelin, 75005 Paris, France}
\affiliation{IPGG, 6 rue Jean-Calvin, 75005 Paris, France}
\author{Thomas Salez}
\affiliation{Univ. Bordeaux, CNRS, LOMA, UMR 5798, 33405 Talence, France}

\author{Zaicheng Zhang}
\email{zhangzaicheng@buaa.edu.cn}
\affiliation{School of Physics, Beihang University, Beijing 102206, China}

\author{Xingkun Man}
\email{manxk@buaa.edu.cn}
\affiliation{School of Physics, Beihang University, Beijing 102206, China}
\affiliation{Peng Huanwu Collaborative Center for Research and Education, Beihang University, Beijing 100191, China}

\begin{abstract}
	Surface creasing in swelling polymer gels is commonly attributed to compressive strain or interlayer swelling mismatch, with the solvent treated as a passive source of swelling.
	Here, using surface-bound polydimethylsiloxane gels swollen by silicone oils, we show that the polymerization degree of a polymeric solvent acts as a previously unrecognized thermodynamic control parameter for surface creasing.
	Increasing solvent chain length drives a transition from transient creasing to a stable surface.
	We rationalize this behavior with a theory coupling Flory--Huggins mixing thermodynamics to mechanical stability.
	The resulting stability boundary closely captures the observed dependence on solvent polymerization degree and network elasticity.
	Thus, solvent molecularity does not merely change the rate or extent of swelling, but determines whether the instability threshold can be reached.
	Our results reveal how the mixing thermodynamics of polymeric solvents can be exploited to promote or suppress macroscopic surface instabilities.
\end{abstract}
\maketitle

\textit{Introduction.---}
	Surface instabilities in soft materials, manifesting as wrinkles, folds, or creases, are hallmark phenomena of nonlinear soft-matter mechanics~\cite{Ben2025Wrinkles, Li2011Surface, Dervaux2012Mechanical, Lestringant2017Buckling, Li2021Elastic, Wang2023Curvature, Hwang2025Surface}. In swelling gels, these instabilities arise from the coupling between solvent uptake and elastic deformation, whereby volumetric swelling combined with surface attachment generates compressive stresses that are released through surface patterning~\cite{Tanaka1987Mechanical, Guvendiren2009SwellingInduced, Lee2012Prescribed, Bertrand2016Dynamics, Kopecz2025Swelling}. 
	Beyond their fundamental interest in soft matter physics, swelling-induced instabilities are relevant to biological morphogenesis~\cite{Savin2011On, Tallinen2016On, Martinez2022Morphological, Guha2025Unbuckling} and to applications in flexible electronics~\cite{Wen2020Dynamic, Tang2021Multilayered, Wang2020flexible} and microfabricated devices~\cite{Chan2018Folding, Ko2022High, Yan2022Photo, Glover2023Creasing}. Understanding what controls the onset of these instabilities is therefore important not only for mechanics, but also for programming surface morphology in responsive materials.

	Among these phenomena, surface creasing is particularly striking because of its localized, sharply nonlinear, and often subcritical onset~\cite{Gent1999Surface, Hohlfeld2011Unfolding, Hohlfeld2012Scale, Razavi2016Surface, Mora2011Surface, Dervaux2011Shape, Bico2018Elastocapillarity, van2021Pinning,Karpitschka2017Cusp,Yoon2010Nucleation, Chen2012Surface, Hohlfeld2013Coexistence, Chen2014Controlled, Mohanan2022Multiscale}. 
	In soft solids, creases have been extensively explored under mechanical compression, from Biot’s linear stability analysis~\cite{Biot1963Surface} to nonlinear treatments capturing localization, self-contact, and energy minimization~\cite{Hong2009Formation, Cao2012Wrinkling, Hohlfeld2012Scale, Hohlfeld2013Coexistence, Tallinen2013Surface, Zalachas2013Crease, Ciarletta2018Matched, Ciarletta2019Soft}. 
	In swollen gels, their formation is commonly interpreted by mapping solvent uptake onto an effective compressive strain or interlayer swelling mismatch, with instability predicted once a critical threshold is exceeded~\cite{Tanaka1987Mechanical, Trujillo2008Creasing, Lee2012Prescribed}. This framework clarified the roles of mechanics, confinement, and swelling kinetics, and recent simulations have further revealed the nonlinear evolution and bifurcation behavior of swelling-induced surface patterns~\cite{Kang2010Effect, Liu2019Elastocapillary, Webber2024Wrinkling}. 
	Conventional descriptions implicitly treat the solvent primarily as a passive source of swelling, with instability onset governed mainly by geometry, transport, and elasticity~\cite{Beebe2000Functional, Drozdov2016Constitutive,VanZanten2024Unconstrained}.

This passive-solvent assumption becomes questionable when gels swell in oligomeric or polymeric solvents, such as silicone oils or polymer melts~\cite{Melekaslan2003PolymerMelt,Mrozek2011SolventSize}. 
In such situations, the solvent polymerization degree $N_{\rm s}$ modifies the mixing entropy and therefore both the osmotic pressure and its response to local volumetric perturbations. Polymeric solvents may thus affect not only the rate and extent of swelling, but also the mechanical stability of the swollen state. Although previous studies have shown that polymeric solvents can substantially alter swelling and compression dynamics~\cite{Aangenendt2020Nonmonotonic,Punter2020Compression,Flapper2023Reversal}, whether solvent molecularity can
control the onset of the surface instability has not been established.

Here, we address this question by combining experiments and theory for surface-bound polydimethylsiloxane (PDMS) gels swollen by silicone oils. By varying the solvent polymerization degree $N_{\rm s}$ and the gel shear modulus $G$, we construct a phase diagram separating creased and stable surfaces. 
The surface response changes systematically with $N_{\rm s}$ despite comparable normalized swelling kinetics and cumulative solvent-uptake profiles. A model coupling Flory--Huggins mixing thermodynamics to mechanical
stability relates this transition to the competition between the swelling ratio at equilibrium $\lambda_{\rm eq}$ and the critical swelling ratio $\lambda_{\rm cri}$ for the onset of instability. The model therefore shows that solvent chain length shifts the mechanical instability threshold through its effect on the osmotic response,  rather than merely rescaling the swelling kinetics or reducing the equilibrium uptake.

\begin{figure}
    \centering
    \includegraphics[width=0.85\linewidth]{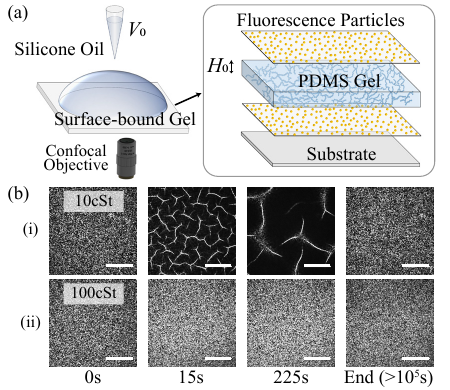}
    \caption{Swelling-induced surface responses of a surface-bound gel.
    (a) Schematic of the experimental setup. A droplet of silicone oil with volume $V_0= 50~\mu\mathrm{L}$ is deposited on the free surface of a PDMS gel layer of initial thickness $H_0$, which is bonded to a rigid substrate. Fluorescent nanoparticles are grafted onto the gel surfaces to enable confocal visualization of surface deformation during swelling.
    (b) Time evolution of the gel ($G \simeq 6{\rm kPa}$) surface swollen by silicone oils with different viscosities (polymerization degrees).
    (i) For a low-viscosity oil (10\,cSt), pronounced surface creases rapidly develop and subsequently relax during swelling.
    (ii) For a high-viscosity oil (100\,cSt), the surface remains smooth throughout the swelling process.
    Images are shown at the initial state (0\,s), 15\,s, 225\,s, and at long times ($>10^5$\,s).
    Scale bars: $100~\mu$m.}
    \label{fig1}
\end{figure}

\textit{Experimental Results.---}
We first examine the surface response of surface-bound polymer gels during solvent swelling.
As schematically illustrated in Fig.~\ref{fig1}(a), a thin layer of PDMS gel (Sylgard 184, Dow) with an initial thickness $H_0 \approx 120$--$130~\mu\mathrm{m}$ is firmly bonded to a rigid glass substrate.
A droplet of dimethyl-silicone oil (XIAMETER PMX-200, Dow) with volume $V_0=50~\mu{\rm L}$ is deposited at room temperature ($25^\circ \rm C$) on the free surface, initiating swelling from above.
Fluorescent polystyrene nanoparticles grafted onto the gel surfaces allow for three-dimensional confocal imaging of the evolving surface morphology~\cite{Xu2010Imaging, Smith2021Droplets, Tyagi2022Solute, Yang2024Dehydration, Yang2025Three}. 

Representative surface morphologies are shown in Fig.~\ref{fig1}(b) for silicone oils of different viscosities and fixed shear moduli $G\simeq 6 \rm{kPa}$.
For a low-viscosity oil (i, $10~\mathrm{cSt}$), pronounced surface creases emerge shortly after the onset of swelling, deepen transiently, and eventually relax at long times.
By contrast, for a high-viscosity oil (ii, $100~\mathrm{cSt}$), the gel surface remains smooth throughout the entire swelling process.  
These qualitatively distinct surface responses occur under otherwise identical experimental
conditions, including gel geometry, confinement, and swelling protocol.  
The same qualitative distinction is reproduced in gels without fluorescent-particle labeling (see SI).

\begin{figure}
    \centering
    \includegraphics[width=0.9\linewidth]{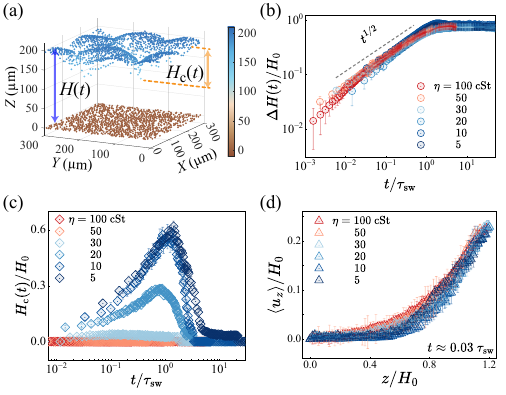}
	\caption{Characteristics of swelling-induced surface instability.
    (a) Three-dimensional reconstruction of the gel surface during swelling, visualized from fluorescent nanoparticles grafted onto the top and bottom surfaces. 
    The instantaneous statistical average gel thickness $H(t)$ and the average crease depth $H_{\rm c}(t)$ are extracted from the reconstructed particle positions.
    (b) Normalized gel thickness evolution $\Delta H(t)/H_0$ as a function of normalized time $t/\tau_{\rm sw}$ for gels swollen by silicone oils of different viscosities $\eta$.
    (c) Temporal evolution of the normalized crease depth $H_{\rm c}/H_0$ as a function of $t/\tau_{\rm sw}$ for gels swollen by silicone oils of different viscosities.
    (d) Normalized vertical-displacement profiles $\langle u_z\rangle/H_0$ versus $z/H_0$ at $t\approx0.03\tau_{\rm sw}$ for different oil viscosities.
}
    \label{fig2}
\end{figure}

To identify the origin of the distinct surface responses in Fig.~\ref{fig1}, we quantify both the swelling kinetics and the evolving surface deformation. From three-dimensional reconstructions of the fluorescent nanoparticle positions shown in Fig.~\ref{fig2}(a), we extract the instantaneous average gel thickness over the \textit{x-y} plane, $H(t)$, together with the average crease depth, $H_{\rm c}(t)$.
A sample is classified as creased when $\max_t H_{\rm c}(t)>H_{{\rm c},\rm th} =3\delta H_0\simeq3~\mu\mathrm{m}$, where $\delta H_0$ is the initial film-thickness uncertainty (see SI).
Figure~\ref{fig2}(b) shows the normalized thickness increase $\Delta H(t)/H_0=(H(t)-H_0)/H_0$ versus the rescaled time $t/\tau_{\rm sw}$, where the characteristic swelling time $\tau_{\rm sw}$ is the time required to reach $90\%$ of the final thickness. Strikingly, all datasets collapse onto a single master curve, with an early-time diffusion-like scaling $\Delta H(t)/H_0\sim (t/\tau_{\rm sw})^{1/2}$ prior to saturation. 
This collapse demonstrates that the normalized swelling kinetics are comparable across all solvent conditions, despite their different absolute swelling times.

The surface response, however, is not universal. Figure~\ref{fig2}(c) shows that the normalized crease depth $H_{\rm c}/H_0$ undergoes pronounced transient growth for low-viscosity oils and is progressively suppressed with increasing viscosity. 
For the 6-kPa gel, the 50-cSt oil lies near the experimental stability boundary and yields both creased and stable outcomes in repeated experiments, whereas the surface consistently remains stable with the 100-cSt oil.

To probe swelling across the film thickness, we track quantum dots dispersed in the gel and obtain the height-resolved vertical displacement $\langle u_z\rangle(z,t)$. Under one-dimensional confined swelling, this displacement measures the cumulative solvent uptake per unit area from the bonded substrate up to height $z$ (see SI for details). At $t\approx0.03\tau_{\rm sw}$, the normalized profiles $\langle u_z\rangle/H_0$ remain comparable across the oils [Fig.~\ref{fig2}(d)] despite their different surface responses. These observations suggest that macroscopic swelling transport alone is insufficient to explain the creased-to-stable transition.

Gel permeation chromatography measurements show that the number-average polymerization $N_{\rm s}$ of the silicone oils increases monotonically with their viscosity (see Figs.~S3 and S4), providing the basis for using $N_{\rm s}$ as the solvent control parameter in the theory~\cite{Rubinstein2003Polymer}.
This identifies the solvent molecular architecture as the potential candidate for the missing control parameter and motivates the theoretical analysis below. Next, we show how $N_{\rm s}$ modifies the thermodynamic driving for swelling and, through it, the mechanical stability of the swollen state.

\textit{Minimal Theory.---} 

To rationalize the observed dependence on solvent polymerization degree, we model the swollen gel by coupling network elasticity with
Flory--Huggins mixing thermodynamics. The solvent
polymerization degree $N_{\rm s}$ enters the mixing free energy
and thereby the osmotic response of the gel. We then determine
separately the \emph{equilibrium} swelling ratio $\lambda_{\rm eq}$
and the \emph{critical} swelling ratio $\lambda_{\rm cri}$ for surface
creasing, whose competition sets whether creasing is
accessible during swelling.

% (free energy: neo-Hookean + Flory--Huggins)
We describe the swollen gel by a continuum free-energy density $f$ (per unit dry polymer volume), written as the sum of an elastic contribution and a network-solvent mixing contribution, $f=f_{\rm el}+f_{\rm mix}$.
Surface energy is neglected at leading order since the estimated elastocapillary length, $\ell_{\rm ec}\simeq0.06$--$1.1~\mu\mathrm{m}$, is much smaller than the gel thickness $H_0$ (see SI)~\cite{Style2017Elastocapillarity,Bico2018Elastocapillarity,Aubouy2000Interfacial}.

For the network elasticity, we adopt a neo-Hookean form, $f_{\rm el}({\bf F}, \phi) = \frac{G}{2}\left[ \mathrm{tr}\!\left({\bf F}^{\!\top}{\bf F}\right) - 3 -2\ln (\det (\bf F)) \right]$
where $\mathbf{F}$ is the deformation gradient, $G=nk_{\rm B}T$ is the network shear modulus with $k_{\rm B}T$ the thermal energy and $n$ the number density of strands~\cite{Biot1963Surface, Kang2010Effect}.
The mixing free energy is taken in the Flory--Huggins form,
$f_{\rm mix}=\frac{1}{1-\phi}\frac{k_{\rm B}T}{v} \left[\frac{1}{N_{\rm s}}\,\phi\ln\phi + \chi\,\phi(1-\phi) \right]$, where $\phi$ is the solvent volume fraction,  $v$ is the effective reference volume, and $\chi$ is the Flory--Huggins interaction parameter~\cite{Rubinstein2003Polymer, Kang2010Effect}. 
Importantly, $N_{\rm s}$ and $\chi$ represent distinct thermodynamic controls: $N_{\rm s}$ governs the chain-translational entropy, whereas $\chi$ characterizes polymer--solvent interactions.
For the chemically matched PDMS network/silicone-oil system, we use the athermal reference $\chi=0$, avoiding an additional adjustable interaction parameter (see SI).
Using incompressibility, $\phi=1-J^{-1}$ with the volumetric swelling ratio $J=\det\mathbf{F}$, so that the mixing free energy can be expressed directly in terms of the volumetric swelling ratio $J$ (see SI). 

% osmotic pressure
The osmotic pressure driving solvent uptake is obtained from the mixing free energy as
$\Pi\equiv-\partial f_{\rm mix}/\partial J$. For a surface-bound gel swelling uniaxially, $\mathbf{F}=\mathrm{diag}(1,1,\lambda_h)$ and the volumetric swelling ratio $J$ reduces to the swelling ratio $\lambda_{\rm h}$ in the $z$-direction \textit{i.e.}, $J=\lambda_{\rm h}$. 
The Flory--Huggins mixing free energy $f_{\rm mix}$ then gives the dimensionless $N_{\rm{s}}$-dependent osmotic pressure in the form of
\begin{equation}
    \widehat{\Pi}(\lambda_{\rm h}) =- \frac{1}{N_{\rm s}}\!\left[ \ln\!\left(1-\frac{1}{\lambda_{\rm h}}\right)+\frac{1}{\lambda_{\rm h}}\right]
    - \frac{\chi}{\lambda_{\rm h}^{2}},
    \label{Pi}
\end{equation}
where $\widehat{\Pi}=\Pi/K_{\rm m}$ and $K_{\rm m}=k_{\rm B}T/v$ is the effective mixing-energy scale.

%% calculation for lambda_cri

To estimate the onset of surface instability, we perform a linear stability analysis (LSA) about the homogeneously swollen state. A small perturbation is superimposed on this base state, and the constitutive relation and mechanical equilibrium are linearized. The resulting neutral-stability condition defines a critical swelling ratio $\lambda_{\rm cri}$. Because creasing is intrinsically nonlinear and subcritical, $\lambda_{\rm cri}$ should be interpreted as a linear neutral-stability threshold rather than the exact nucleation threshold of a finite-amplitude crease~\cite{Ciarletta2018Matched,Ciarletta2019Soft} (see SI).
In the short-wavelength limit, $kH_0 \gg 1$, where $k$ is the Fourier wave number, the neutral-stability condition reduces to
\begin{equation}
\left(\lambda_{\rm cri}+\frac{1}{\lambda_{\rm cri}}\right)^2
=
4\lambda_{\rm cri}
\sqrt{
\frac{2+\lambda_{\rm cri}\widehat{K}/\widehat{G}}
{\lambda_{\rm cri}^2+1+\lambda_{\rm cri}\widehat{K}/\widehat{G}}
},
\label{eq:lambda_cri}
\end{equation}
where $\widehat{K}=K/K_{\rm m}$ is the normalized osmotic bulk modulus with $K\equiv-\,{\rm d}\Pi/{\rm d}(\ln J)$
and $\widehat{G}=G/K_{\rm m}=nv$ is the normalized network shear modulus. From Eq.~(\ref{Pi}), the normalized osmotic bulk modulus  $\widehat{K}$  is expressed as:
\begin{equation}
   \widehat{K}(\lambda_{\rm h})=
   \frac{1}{N_{\rm s}} \left( \frac{1}{\lambda_{\rm h}-1} - \frac{1}{\lambda_{\rm h}}\right)
   -\frac{2\chi}{\lambda_{\rm h}^2}.
   \label{K}
\end{equation}
Thus, $N_{\rm s}$ enters the stability problem through the osmotic bulk modulus $\widehat K$, which characterizes the response of the osmotic pressure to a local volumetric perturbation.

\begin{figure}
    \centering
    \includegraphics[width=0.9\linewidth]{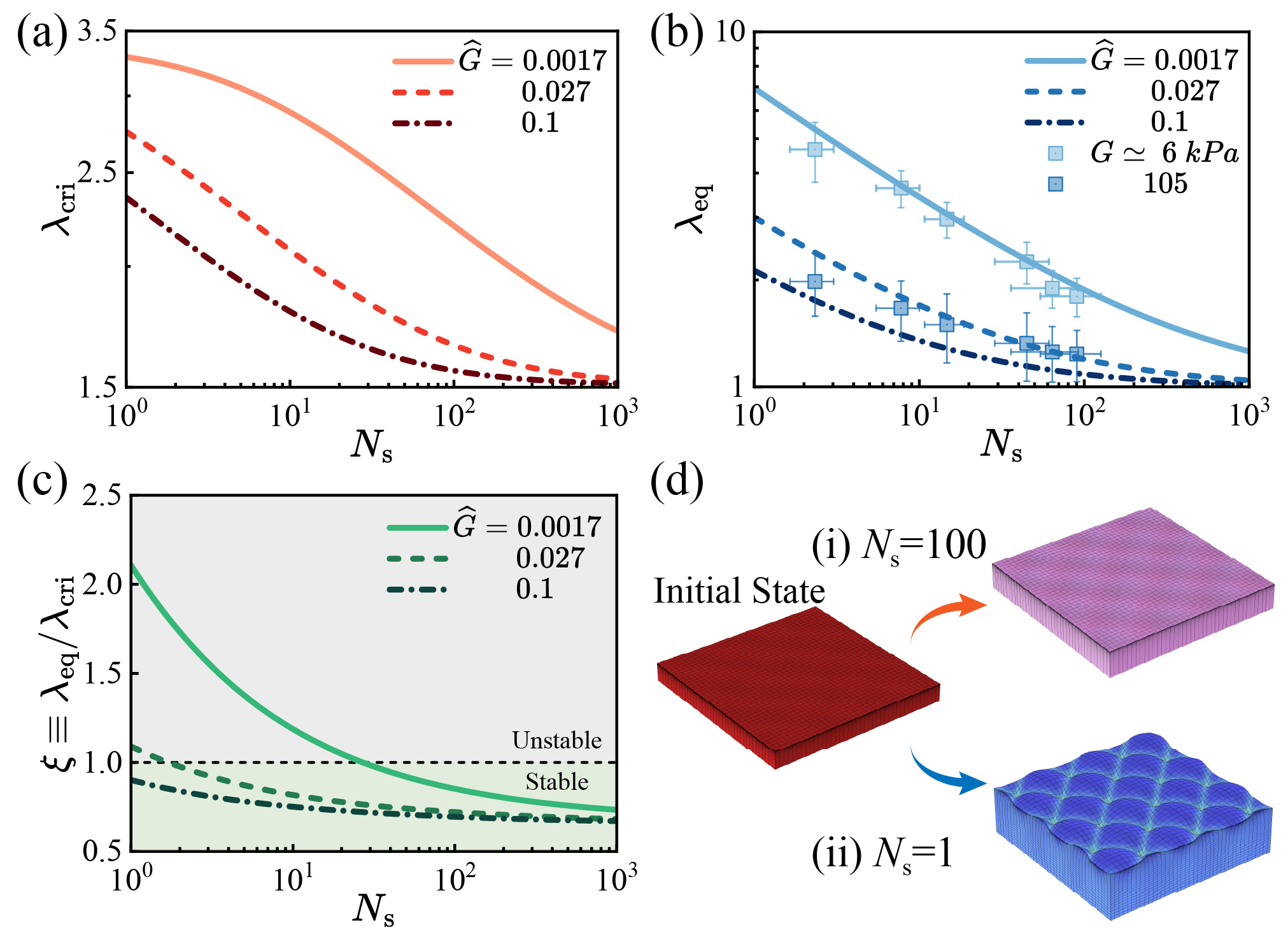}
    \caption{Mechanism governing swelling-induced surface creasing.
    (a) Critical swelling ratio for the onset of surface creasing, $\lambda_{\rm cri}$, plotted as a function of solvent polymerization degree $N_{\rm s}$ for different normalized shear moduli $\widehat{G}$.
    (b) Corresponding equilibrium swelling ratio $\lambda_{\rm eq}$ as a function of $N_{\rm s}$ for the same values of $\widehat{G}$. Symbols denote experimentally measured $\lambda_{\rm eq}$ for gels  with $G\simeq6\,\rm{kPa}$ and $G\simeq105\,\rm{kPa}$ swollen by silicone oils with different $N_{\rm s}$.
    (c) Stability index $\xi=\lambda_{\rm eq}/\lambda_{\rm cri}$ as a function of $N_{\rm s}$, with unstable ($\xi>1$) and stable ($\xi<1$) regimes indicated.
    (d) FEM simulations of swelling evolution in a surface-bound gel with a small initial surface perturbation for $N_{\rm s}=100$ and $N_{\rm s}=1$. Here we have set $\chi=0.0$.}
    \label{fig3}
\end{figure}

% Discribe Fig3
Figure~\ref{fig3}(a)  shows the critical swelling ratio
$\lambda_{\rm cri}$ as a function of $N_{\rm s}$ for different
values of $\widehat G$. At fixed $\widehat G$,
$\lambda_{\rm cri}$ decreases with increasing $N_{\rm s}$,
while at fixed $N_{\rm s}$ it decreases with increasing
$\widehat G$. 
Taken alone, these trends of $\lambda_{\rm cri}$ would imply that increasing either $N_{\rm s}$ or $\widehat G$ makes the surface more susceptible to instability, in apparent contradiction with the experiments.
This indicates that the instability threshold alone is
insufficient to determine whether creasing occurs during
swelling.

% elastic stress
Whether this instability threshold is reached, however, depends on the equilibrium swelling ratio $\lambda_{\rm eq}$. For confined uniaxial swelling, $\lambda_{\rm eq}$ is set by the balance between the elastic restoring stress $\sigma^{\rm el}_{zz}$ of the network and the mixing osmotic pressure $\Pi$. For the neo-Hookean network, the normalized elastic stress $\widehat{\sigma}^{\rm el}_{zz}=\sigma^{\rm el}_{zz}/K_{\rm m}$ is
\begin{equation}
	\widehat{\sigma}^{\rm el}_{zz}
	=
	\widehat{G}
	\left(
	\lambda_{\rm h}-\frac{1}{\lambda_{\rm h}}
	\right).
	\label{sigma_zz}
\end{equation}
In normalized form, the mechanical equilibrium condition is
\begin{equation}
	\widehat{\sigma}^{\rm el}_{zz}(\lambda_{\rm eq})
	=
	\widehat{\Pi}(\lambda_{\rm eq}).
	\label{eq:lambda_eq_balance}
\end{equation}
This balance is equivalent to minimizing the total free-energy density with respect to $\lambda_{\rm h}$ (see SI).

%% figure for the therotical model

Figure~\ref{fig3}(b) shows $\lambda_{\rm eq}$ obtained from Eq.~\eqref{eq:lambda_eq_balance} as a function of $N_{\rm s}$ for different values of $\widehat{G}$.
At fixed $\widehat{G}$, $\lambda_{\rm eq}$ decreases monotonically with increasing $N_{\rm s}$, reflecting the suppression of mixing entropy for polymeric solvents.
Increasing $\widehat{G}$ further reduces $\lambda_{\rm eq}$ by enhancing the elastic resistance to swelling.
The symbols in Fig.~\ref{fig3}(b) denote experimentally measured equilibrium swelling ratios for gels with shear moduli $G\simeq6\,\mathrm{kPa}$ and $G\simeq105\,\mathrm{kPa}$.
The experimental data are in quantitative agreement with the theoretical predictions using $K_{\rm m}$ as the only fitting parameter, yielding a characteristic scale $K_{\rm m}\simeq4~\mathrm{MPa}$.

Creasing can occur only if the instability threshold is reached before swelling equilibrates, \textit{i.e.}, when $\lambda_{\rm eq}>\lambda_{\rm cri}$. This competition is captured by the dimensionless stability parameter
\begin{equation}
\xi \equiv \frac{\lambda_{\rm eq}}{\lambda_{\rm cri}}.
\label{eq:stability_index_def}
\end{equation}
Figure~\ref{fig3}(c) shows $\xi$ as a function of $N_{\rm s}$.
For $\xi>1$, the gel reaches the instability threshold $\lambda_{\rm cri}$ before attaining thermodynamic equilibrium, so that surface creasing can occur during swelling. By contrast, when $\xi<1$, swelling reaches equilibrium at $\lambda_{\rm eq}$ without ever crossing the instability threshold, and the surface remains stable.  
The stability parameter $\xi$ therefore indicates whether the instability threshold is accessible, but does not describe the subsequent crease dynamics.
Figure~\ref{fig3}(c) further shows that increasing $N_{\rm s}$ or $\widehat{G}$ reduces $\xi$, indicating enhanced surface stability, in agreement with the experiments. This resolves the apparent contradiction in Fig.~\ref{fig3}(a): although both $\lambda_{\rm cri}$ and
$\lambda_{\rm eq}$ decrease as either $N_{\rm s}$ or $\widehat{G}$ increases, $\lambda_{\rm eq}$ decreases more rapidly, driving $\xi$ toward the stable regime.

\begin{figure}[b]
	\centering
	\includegraphics[width=0.8\linewidth]{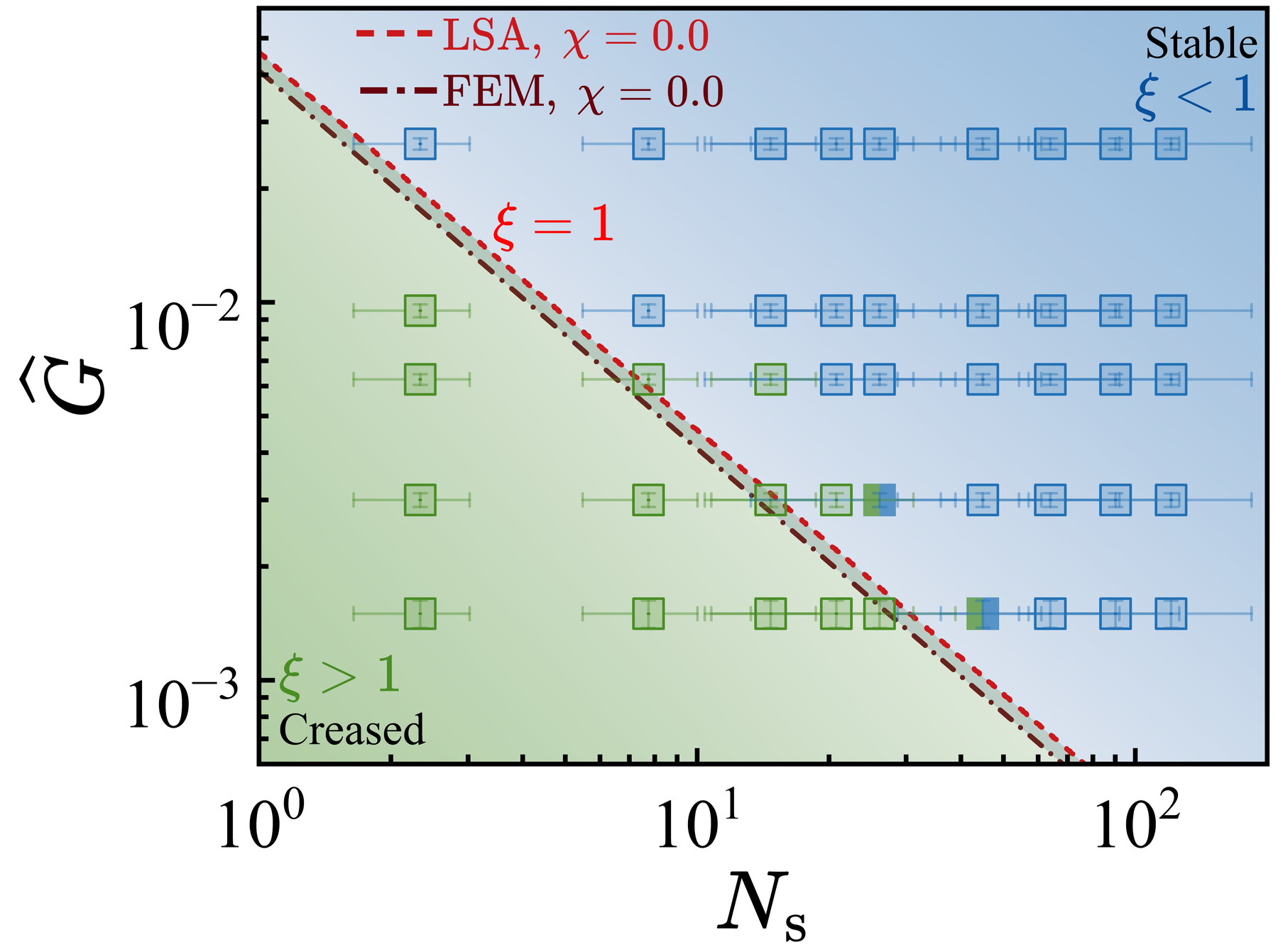}
	\caption{
		Experimental phase diagram and theoretical stability boundaries
		for swelling-induced surface creasing. Symbols denote
		experimental outcomes in the $(N_{\rm s},\widehat{G})$ plane,
		classified as creased (green) or stable (blue). Half-filled
		symbols indicate mixed creased/stable outcomes in repeated
		experiments. Error bars indicate the experimental uncertainties
		in $N_{\rm s}$ and $\widehat{G}$. The red dashed and dark red
		dash-dotted lines denote the LSA ($\xi=1$) and FEM stability
		boundaries, respectively, both evaluated with $\chi=0$. Shaded
		regions indicate the predicted stable ($\xi<1$) and unstable
		($\xi>1$) regimes.
	}
	\label{fig4}
\end{figure}

Time-dependent finite-element (FEM) simulations with a small initial surface perturbation ($\sim10^{-3}H_0$) further illustrate the above criterion (see SI). For $N_{\rm s}=100$ ($\xi<1$), the perturbation decays and the surface remains smooth [Fig.~\ref{fig3}(d.i)], whereas for $N_{\rm s}=1$ ($\xi>1$), the perturbation grows and develops into a pronounced surface crease [Fig.~\ref{fig3}(d.ii)]. The subsequent relaxation and disappearance of the experimental creases lie outside the present model and may involve finite-volume solvent redistribution and nonlinear relaxation.

Guided by the theory, we varied the gel shear modulus $G$
through the PDMS crosslinking density and examined silicone
oils spanning different values of $N_{\rm s}$, with $G$
measured independently by rotational rheometry.
Figure~\ref{fig4} shows the resulting experimental phase
diagram in the $(N_{\rm s},\widehat{G})$ plane, where
$\widehat{G}=G/K_{\rm m}$ and $K_{\rm m}$ is obtained from
the equilibrium-swelling fits in Fig.~\ref{fig3}(b). For a
given gel modulus, increasing $N_{\rm s}$ suppresses surface
creasing, whereas for a fixed solvent, increasing $G$
stabilizes the surface.
Additional experiments with gels swollen by a homologous series of $n$-alkanes show that increasing alkane chain length similarly stabilizes the surface, suggesting that the effect is not specific to silicone oils (see SI).

The LSA boundary $\xi=1$, evaluated with $\chi=0$, captures the main experimental trends with both $N_{\rm s}$ and $\widehat{G}$. The corresponding FEM boundary, which accounts for finite deformation and nonlinear evolution toward surface self-contact, lies close to the LSA prediction and exhibits the same parameter dependence. Together, these results support $\xi=1$ as a minimal criterion for determining whether the instability threshold can be reached, while residual quantitative deviations likely reflect physical effects beyond the present model. 

\textit{Conclusion.---}
In summary, we identify the polymerization degree $N_{\rm s}$ of a polymeric solvent as a thermodynamic control parameter for swelling-induced surface creasing. 
The transition from creased to stable surfaces cannot be inferred from swelling kinetics or equilibrium uptake alone. By modifying the osmotic
response, $N_{\rm s}$ shifts both the equilibrium swelling ratio $\lambda_{\rm eq}$ and the critical swelling ratio $\lambda_{\rm cri}$ for mechanical instability, so that their competition determines whether the instability threshold can be reached. This result links molecular-scale mixing entropy to macroscopic surface stability. 
Solvent chain length should therefore be considered alongside network elasticity when assessing the stability of swollen soft materials.

\textit{Acknowledgment.---}
We thank Masao Doi, Shiyuan Hu, and Robert Style for fruitful discussions, and Zhiwei Liu for the help with the rheometer measurements. This work was supported in part by the National Natural Science Foundation of China (No. 22473005 and No. 12404234), the Beijing Natural Science Foundation (No. 2262016), the Fundamental Research Fund for Central Universities (No. GW2025-33) and the Open Fund of the State Key Laboratory of Nonlinear Mechanics. We also acknowledge financial support from the European Union through the European Research Council under EMetBrown (ERC-CoG-101039103) grant, as well as from the Agence Nationale de la Recherche under EMetBrown (ANR-21-ERCC-0010-01), Softer (ANR21-CE06-0029), Fricolas (ANR-21-CE06-0039) and MS-DOS (ANR-25-CE06-3953-02) grants, and the Interdisciplinary and Exploratory Research program under MISTIC grant at the University of Bordeaux, France. Besides, we acknowledge the support from the R\'eseaux de Recherche Impulsion Frontiers of Life which received financial support from the French government in the framework of the University of Bordeaux's France 2030 program. Finally, we thank the Soft Matter Collaborative Research Unit, Frontier Research Center for Advanced Material and Life Science, Faculty of Advanced Life Science at Hokkaido University, Sapporo, Japan, and the CNRS International Research Network between France and India on Hydrodynamics at small scales: From soft matter to bioengineering.

\bibliography{references.bib}
\end{document}